\documentclass[aps]{JHEP3}

\newcommand{\ep}{\epsilon}

\newcommand{\vphi}{\varphi}

\newcommand{\pa}{\partial}
\newcommand{\td}{\tilde}
\newcommand{\sla}[1]{\slash\!\!\! #1}
\newcommand{\Sla}[1]{\slash\!\!\!\! #1}

\title{New Deformation of ${\cal N}=4$ Super-Yang-Mills Theory}

\author{Xiao-Jun Wang\thanks{This work is partly supported by NSF
of china, 90103002.}\\
Institution of Theoretical Physics, Beijing, 100080, P.R.China \\
E-mail: wangxj@itp.ac.cn}

\abstract{We propose a new algebraic deformation of ${\cal N}=4$
SYM via decomposition of spinor and scalar fields in vector
supermultiplet. This decomposition generates degrees of freedom
of usual quarks and leptons and the deformation model is a low
energy effective model. We show that supersymmetry is broken in
certain limit and the deformation model reduces to a SM-like
model, or a QCD-like model. Meanwhile, gauge symmetry can be
spontaneously broken by nontrivial supersymmetry vacuum.}


\begin{document}

\section{Introduction}

Today, more and more physicists have believed that the
supersymmetry (SUSY), which is treated as maximum symmetry
allowed by S-matrix\cite{CM67}, is fundamental symmetry of action
of quantum field theory\cite{DG81}. During the past two decades,
many SUSY models, such as supergravity, super Yang-Mills theory
(SYM) (and consequently, the supersymmetric version of standard
model (SSM)) have been widely studied. All of these SUSY models
include large number of field degrees of freedom, and most of
them are out of our current observable world. For instance, in
SSM, besides of well-known standard model particles: gluon, quark,
lepton, ..., there are also their superpartners: gluino, squark,
slepton, .... However, all of these superpartners were not found
by current experiment. Traditionally, it is believed that those
missed degrees of freedom are very heavy when SUSY is
spontaneously broken. So that they did not be observed in current
experiment spectrum. Unfortunately, so far we still can not find
a simple and wide-accepted mechanism to spontaneously break SUSY
even though large number of remarkable results have been
achieved\cite{CAN82,DN93,Extra}. Consequently, we still can not
satisfactorily interpret how these particles are missed. This may
be because the mechanism of SUSY spontaneously breaking is very
complicated, but are there other possibilities? This is just
purpose of this paper.

Our work was motivated by Maldacena's conjecture on Ads/CFT
corresponds\cite{Maldacena98}, which states that ${\cal N}=4$ SYM
in four dimension is dual to IIB string theory on $Ads_5\times
S_5$. This conjecture provides a principle method to deal with
strong coupling case of quantum field theory. In particular, it
seem to emerge a possibility, that we can directly calculate
nonperturbative effect of QCD from QCD. This purpose has been
partly achieved in some literature\cite{Polchinski}. For example,
the spectrum of glueball has been calculated\cite{glueball} and
agree with result of lattice QCD. However, it is still very
difficult to deal with interaction of state involving quarks. The
essential reason is that, in ${\cal N}=4$ SYM, fermion fields are
adjoint representation of gauge group rather than fundamental
representation of gauge group. The usual method to construct SSM
from SYM is to add extra chiral supermultiplets, which are
fundamental representation of gauge group and represent quark,
lepton and their superpartners. Unfortunately, we can not obtain
this type of chiral superfields from string theory and its
dimensional reduction. It seem to imply that the quark and lepton
\footnote{In this paper, quark and lepton denote spin $1/2$
fermion which belong to fundamental representation of gauge
group.} can not be defined (perturbatively at least) in string
theory. However, whether can quark or lepton be generated by
vector multiplet itslef via deformation of SYM or spontaneously
breaking of SYM? We will suggest a new deformation of ${\cal
N}=4$ SYM in this paper and show that this can indeed be achieved.

Our idea is originated by knowledge of group theory: The adjoint
representation of $SU(N)$ group can be generated by directly
multiplication of $N$ and $\bar{N}$ representation of $SU(N)$
group. Then vector spuermultiplet of ${\cal N}=4$ SYM can be
decomposed into a scalar field, four spin $1/2$ fermions and a
vector field. Here all of these scalar field and fermions are
fundamental representation of gauge group. This idea proposes a
simple algebra deformation on ${\cal N}=4$ SYM and the deformation
indeed yields similar field degrees of freedom of SM-like model
or of QCD-like model. R-symmetry, furthermore, will correspond to
generation symmetry of SM-like model, or flavor symmetry of
QCD-like model. The simple dimensional analysis requires a scale
parameter $M$ appearing in deformation model. It indicates that
the deformation model will be low energy effective model, and its
Lagrangian will be expanded in power of $1/M^2$. When energy is
much lower than scale $M$, only part of Lagrangian is survived.
Thus this method provides a natural mechanism to ``spontaneously
break'' SUSY. Meanwhile, the new scalar fields in deformed model
are no longer Higgs bosons, rather, they are Goldstone bosons
corresponding to SUSY spontaneously breaking, or even they do not
appear as dynamical degrees of freedom in certain condition.

In the next section, we propose a simple algebra deformation on
${\cal N}=4$ SYM. It yields a low energy effective theory. In
section 3, We discuss some properties of this deformation model.
Section 4 is devoted a brief summary and discussion.

\section{New Deformation of ${\cal N}=4$ SYM}

The supermultiplets of ${\cal N}=4$ super Yang-Mills theory
contains a spin-1 gauge boson $A_\mu^a$, four spin $1/2$ Majorana
spinors $\Psi_A^a$ and six scalar fields $\Phi_{AB}^a$ (they
belong to an antisymmetric tensor representation of $SU(4)$ group
so that $A,\;B$ are antisymmetric). Here $A,\;B=1,2,3,4$ denote
indexes of $SU(4)$ R-symmetry, $a,\;b=1,...,N^2-1$ denote indexes
of adjoint representation of gauge group. The Lagrangian density
of ${\cal N}=4$ SYM can compactly  be written in the manifestly
$SU(4)$-invaraint form
\begin{eqnarray}\label{01}
{\cal L}&=&\frac{1}{2}\sum_{AB}Tr\{(D_\mu
\Phi_{AB})(D^\mu\Phi_{AB})^{\dag}\}+i\sum_{A}Tr\left(\bar{\Psi}_A\;
{\Sla D}\Psi_A\right) \nonumber \\
&&-\sqrt{2}{\rm Re}\;\sum_{AB}Tr\left(\bar{\Psi}_A\gamma_5
[\Phi_{AB},\Psi_B]\right)
-\frac{1}{8}\sum_{ABCD}Tr\left([\Phi_{AB},\Phi_{CD}]^2\right)
  \nonumber \\
&&-\frac{1}{4}F_{\mu\nu}^aF^{a\mu\nu}
+\frac{g^2\theta}{64\pi^2}\ep^{\mu\nu\rho\sigma}F_{\mu\nu}^a
F_{\rho\sigma}^a,
\end{eqnarray}
where
\begin{eqnarray}\label{02}
D_\mu\Psi_A&=&\pa_\mu\Psi_A+i[A_\mu,\Psi_A],\nonumber \\
D_\mu\Phi_{AB}&=&\pa_\mu\Phi_{AB}+i[A_\mu,\Phi_{AB}].
\end{eqnarray}
The Majorana spinor fields and scalar fields in the vector
supermultiplets can be decomposed to a set of scalar fields
$\vphi_i$ and four sets of Majorana spinors $\lambda_{Ai}$,
\begin{eqnarray}\label{03}
\Psi_{Aij}&=&\frac{1}{\sqrt{2}}(\phi_i\lambda_{Aj}
 +\lambda_{Ai}\phi_j
 -\frac{2}{N}\phi_l\lambda_{Al}\delta_{ij}), \nonumber \\
\bar{\Psi}_{Aij}&=&\frac{1}{\sqrt{2}}(\phi_i\bar{\lambda}_{Aj}
+\bar{\lambda}_{Ai}\phi_j
 -\frac{2}{N}\phi_l\bar{\lambda}_{Al}\delta_{ij}),  \\
(\Phi_{AB})_{ij}&=&M^{-2}(\bar{\lambda}_{Aj}\gamma_5\lambda_{Bi}
 -\bar{\lambda}_{Bj}\gamma_5\lambda_{Ai})
 -\frac{1}{2}Z_{AB}U_{ij}(\vphi)-\frac{1}{2}(1-\frac{2}{N})
 Z_{AB}\delta_{ij}, \nonumber \\
\end{eqnarray}
where $i,\;j=1,...,N$ are indexes of gauge group,
\begin{eqnarray}\label{04}
\phi_i=\frac{\vphi_i}{\sqrt{\sum_l\vphi_l^2}}\quad
&\Longrightarrow &\quad \sum_i\phi_i^2=1, \nonumber \\
U_{ij}=\delta_{ij}-2\phi_i\phi_j\quad &\Longrightarrow &\quad
U^{-1}=U,
\end{eqnarray}
and $Z_{AB}$ is central charge associating with extended
supersymmetric algebra
\begin{eqnarray}\label{05}
\{Q_{Aa},Q_{Bb}\}=e_{ab}Z_{AB}\hspace{0.5in}
e=i\sigma_2=\left({{0,\quad 1}\atop {-1,\quad 0}}\right).
\end{eqnarray}

Since $\lambda_A$ are fundamental representation of $SU(N)$ gauge
group, its gauge transformation $\lambda_A\rightarrow \lambda'_A=
\Omega(x)\lambda_A (\Omega^{\dag}\Omega=1)$ makes $\lambda_{Ai}$
no longer be Majorana spinor in general gauge. In this paper, it
is convenient to adopt a specific version of Wess-Zumino gauge
(hereafter we call it as SWZ gauge) in which $A_\mu$ is
antisymmetric, i.e., its nonzero elements form an adjoint
representation of $SO(N)$ subgroup of $SU(N)$ group. In SWZ gauge,
$\lambda_A$ will be Majorana spinor, and the covariant derivative
\begin{eqnarray}\label{06}
D_\mu\lambda_A&=&\pa_\mu\lambda_A+iA_\mu\lambda_A,\nonumber \\
D_\mu\bar{\lambda}_A&=&\pa_\mu\bar{\lambda}_A-i\bar{\lambda}_A
A_\mu
\end{eqnarray}
are well-defined. In addition, since $\Psi_A$ is Majorana spinor,
eq.~(\ref{03}) requires that $\phi$ is real field in SWZ gauge.

Now inserting eq.~(\ref{03}) into Lagrangian of ${\cal N}=4$ SYM,
we can derive effective Lagrangian of  deformation model.
\begin{enumerate}
\item Using the properties of Majorana spinors,
$\bar{\lambda}_{Ai}\gamma_\mu\lambda_{Aj}=-\bar{\lambda}_{Aj}
\gamma_\mu\lambda_{Ai}$, and defining ${\cal
M}_{ij}=\phi_i\phi_j$, we have ${\bar\lambda}_A{\cal
M}\Sla{A}{\cal M}\lambda_A={\bar\lambda}_A(\sla{\pa}{\cal M})
\lambda_A={\bar\lambda}_A{\cal M}(\sla{\pa}{\cal M}){\cal
M}\lambda_A=0$. So that
\begin{eqnarray}\label{07}
Tr\left(\bar{\Psi}_A\;{\Sla D}\Psi_A\right)
&=&\bar{\lambda}_A\;{\Sla D}\lambda_A+(\frac{1}{2}+\frac{1}{N})
\bar{\lambda}_A\gamma_\mu(\pa^\mu {\cal M}{\cal M}- {\cal
M}\pa^\mu{\cal M})\lambda_A \nonumber \\
&&+(1-\frac{2}{N})\bar{\lambda}_A{\cal M}(\sla{\pa}\lambda_A)+
\bar{\lambda}_A({\cal M}\Sla{A}+\Sla{A}{\cal M})\lambda_A.
\end{eqnarray}
The kinetic term of spinor $\lambda_A$ can be diagonalized via
field redefinition
\begin{eqnarray}\label{08}
\lambda_A\;\longrightarrow\; (1+a{\cal M})\lambda_A,
\end{eqnarray}
with $a=-1+\sqrt{\frac{N}{2(N-1)}}$. Then eq.~(\ref{07}) becomes
\begin{eqnarray}\label{09}
Tr\left(\bar{\Psi}_A\;{\Sla
D}\Psi_A\right)&=&\bar{\lambda}_A\;{\Sla D}\lambda_A+\frac{1}{2}
(1+2a)\bar{\lambda}_A\gamma_\mu(D^\mu U^{\dag}U- U^{\dag} D^\mu U)
\lambda_A \nonumber \\
&=&\bar{\lambda}_A\;{\Sla D}\lambda_A-
(1+2a)\bar{\lambda}_A\gamma_\mu (U^{\dag} D^\mu U)\lambda_A,
\end{eqnarray}
where
\begin{eqnarray}\label{10}
D_\mu U=\pa_\mu U+iA_\mu U-iUA_\mu .
\end{eqnarray}
Furthermore, in order to make the coupling between spinor and
gauge boson be standard form, we can let that
\begin{eqnarray}\label{11}
A_\mu=A'_\mu+ibU^{\dag}D'_\mu U,\hspace{0.5in}
b=\frac{1}{2}\left(-1+\sqrt{\frac{N-1}{2N}}\right),
\end{eqnarray}
where
\begin{eqnarray}\label{12}
D'_\mu U=\pa_\mu U+iA'_\mu U-iUA'_\mu .
\end{eqnarray}
Then finally we have
\begin{eqnarray}\label{13}
Tr\left(\bar{\Psi}_A\;{\Sla
D}\Psi_A\right)=\bar{\lambda}_A\;{\Sla D}'\lambda_A
\end{eqnarray}
\item Inserting eq.~(\ref{03}) into $F_{\mu\nu}^aF^{a\mu\nu}$ and
$\ep^{\mu\nu\rho\sigma}F_{\mu\nu}^a F_{\rho\sigma}^a$, and using
eq.~(\ref{11}), we have
\begin{eqnarray}\label{14}
Tr\left(F_{\mu\nu}F^{\mu\nu}\right)&=&(1+2b+2b^2)
Tr\left(F'_{\mu\nu}F'^{\mu\nu}\right)
-2b(1+b)Tr\left(F'_{\mu\nu}U^{\dag}F'^{\mu\nu}U\right)
    \nonumber \\
&&-2b^2Tr\left(D'_\mu U^{\dag}D'_\nu UD'^{\mu}U^{\dag}D'^{\nu}U
  -D'_\mu U^{\dag}D'^{\mu} UD'_{\nu}U^{\dag}D'^{\nu}U\right)
    \nonumber \\
&&+4ibTr\left(F'_{\mu\nu}D'^{\mu}U^{\dag}D'^\mu U\right) \\
\ep^{\mu\nu\rho\sigma}Tr\left(F_{\mu\nu}F_{\rho\sigma}\right)&=&
(1-2b+10b^2)\ep^{\mu\nu\rho\sigma}Tr\left(F'_{\mu\nu}F'_{\rho\sigma}
\right) \nonumber \\
&&+2b(1-5b)\ep^{\mu\nu\rho\sigma}Tr\left(F'_{\mu\nu}U^{\dag}
F'_{\rho\sigma}U\right) \nonumber
\end{eqnarray}
\item \mbox{}
\vspace{-0.35in}
\begin{eqnarray}\label{15}
Tr\{(D_\mu \Phi_{AB})(D^\mu\Phi_{AB})^{\dag}\}&=&\frac{f^2}{4}
Tr\left(D'_\mu U^{\dag}D'^{\mu} U\right)+\parbox{1in}{\rm
four fermion} \nonumber \\
&&\parbox{3in}{\rm terms (suppressed by $M^{-4}$)}
\end{eqnarray}
with \parbox{2in}{$$f^2=\sum_{AB}(1+2b)^2|Z_{AB}|^2$$}.
\item The deformation of $\sqrt{2}\{(\Phi_{AB})_{ij}[\bar{\Psi}_{Ajk}
\gamma_5\Psi_{Bki}-\bar{\Psi}_{Aki}\gamma_5\Psi_{Bjk}]\}$ only
yields some four fermion terms which is suppressed by $M^{-2}$.
\item The deformation of the term
$\sum_{ABCD}Tr\left([\Phi_{AB},\Phi_{CD}]^2\right)$ is nothing
other than generating four fermion terms (suppressed by
$Z^2/M^4$), six fermion terms (suppressed by $Z/M^6$) and eight
fermion terms (suppressed by $M^{-8}$).
\end{enumerate}
Finally, the Lagrangian of deformation model is written
\begin{eqnarray}\label{17}
{\cal L}&=&i\sum_A\bar{\lambda}_A\;\Sla{D}\lambda_A -\frac{1}{4}
F_{\mu\nu}^aF^{a\mu\nu}+\frac{g^2\theta}{64\pi^2}
\ep^{\mu\nu\rho\sigma}F_{\mu\nu}^aF_{\rho\sigma}^a
+\frac{f^2}{4} Tr\left(D_\mu U^{\dag}D^{\mu} U\right)\nonumber \\
&&-b(1+b)Tr\left(F_{\mu\nu}F^{\mu\nu}{\cal M}
  -F_{\mu\nu}{\cal M}F^{\mu\nu}{\cal M}\right)
-\frac{i}{2}bTr\left(F_{\mu\nu}D^{\mu}U^{\dag}D^\nu U\right)
\nonumber \\
&&+\frac{b^2}{4}Tr\left(D_\mu U^{\dag}D_\nu
UD^{\mu}U^{\dag}D^{\nu}U-D_\mu U^{\dag}D^{\mu} U
  D_{\nu}U^{\dag}D^{\nu}U\right)     \nonumber \\
&&-\frac{g^2\theta}{8\pi^2}b(1-5b)\ep^{\mu\nu\rho\sigma}
Tr\left(F_{\mu\nu}F_{\rho\sigma}{\cal M}-
  F_{\mu\nu}{\cal M}F_{\rho\sigma}{\cal M}\right) \nonumber \\
&&+\;\parbox{4in}{\rm four fermion terms (suppressed by $M^{-2}$
and $M^{-4}$)}
 \nonumber \\
&&+\;\parbox{4in}{\rm six fermion terms (suppressed by $M^{-6}$)}
   \nonumber \\
&&+\;\parbox{4in}{\rm eight fermion terms (suppressed by
$M^{-8}$)}
\end{eqnarray}

\section{Characterization of The Deformation Model}

In this section we discuss some important properties of the
deformation model parametered by Lagrangian~(\ref{17}).

\subsection*{Gauge invariance}

The Lagrangian~({\ref{17}) is manifest gauge invariant under the
following gauge transformation
\begin{eqnarray}\label{3.1}
\lambda_A\;&\longrightarrow &\;\Omega(x)\lambda_A,\hspace{0.5in}
 \Omega(x)\in SU(N), \nonumber \\
U\;&\longrightarrow &\;\Omega U\Omega^{\dag},\hspace{0.6in} {\cal
M}\;\longrightarrow \;\Omega {\cal M}\Omega^{\dag}, \nonumber \\
A_\mu\;&\longrightarrow &\;\Omega A_\mu\Omega^{\dag}+\frac{i}{g}
\Omega \pa_\mu\Omega^{\dag}.
\end{eqnarray}
In the previous section, the Lagrangian~({\ref{17}) is written in
SWZ gauge. It can be written in general gauge via the above gauge
transformation. In general gauge, however, $\lambda_A$ is neither
Majorana spinor nor Dirac spinor.

\subsection*{Symmetry breaking, Low energy limit}

Two dimensional parameters, $M$ and $Z_{AB}$, present in
Lagrangian~({\ref{17}). So that this deformation model no longer
is a renormalizable theory, rather, it is a low energy effective
model. In other words, it includes the coupling of multi fermions
and is similar to extended Nambu-Jola-Lasinio model. In
Lagrangian~({\ref{17}), although the supersymmetry has become
highly nonlinear, it is still kept. The conformal invariance,
however, has been broken when dimensional parameters are
introduced in deformation~({\ref{03}).

Another interesting issue is low energy limit of this deformation
model, i.e., for energy scale $\mu\ll M$ or $M\to \infty$. In this
limit, the terms on multi fermion coupling can be ignored. Then
the supersymmetry is manifest broken in low energy limit and we
obtain a SM-like model except for the terms associating coupling
of scalar fields. Up to the gauge transformation~{\ref{3.1}),
there are $N$ scalar field in this deformation model. They are
dynamical degrees of freedom of deformation model. However, the
coupling among them fields are nonlinear and their decay constant
$f$ is dimensional. Therefore, these scalar fields are not Higgs
bosons in this SM-like model (for $M\to\infty$), rather, they
become Goldstone bosons corresponding to SUSY breaking.

Since there is another dimensional parameter
$Z=\sqrt{\scriptstyle \sum_{AB}|Z_{AB}|^2}$ in deformation model,
the limit of $M\to \infty$ is not unique. In general, there are
three possible limits:
\renewcommand{\labelenumi}{\roman{enumi}}
\begin{enumerate}
\item $Z\ll\mu\ll M$. From kinetic term of scalar fields we can see
that the physical scalar fields are obtained via field rescaling
$\phi\rightarrow\phi/f\sim \phi/Z$. Then dynamics is dominant by
Goldstone bosons in this case.
\item $Z\sim\mu\ll M$. In this case, both of the interaction of
Goldstone bosons and one of other fields are important.
\item $Z\sim M\gg\mu$. This case is very interesting and important.
All interaction associating Goldstone bosons are frozen out in
this case. Then remain dynamics is parametered by the following
Lagrangian
\begin{eqnarray}\label{3.2}
{\cal L}&=&i\sum_A\bar{\lambda}_A\;\Sla{D}\lambda_A -\frac{1}{4}
F_{\mu\nu}^aF^{a\mu\nu}+\frac{g^2\theta}{64\pi^2}
\ep^{\mu\nu\rho\sigma}F_{\mu\nu}^aF_{\rho\sigma}^a
 \nonumber \\
&&-f^2Tr\left([A_\mu,<{\cal M}>][A^{\mu},<{\cal M}>]\right).
\end{eqnarray}
Here $<{\cal M}>$ denotes expectation value of ${\cal M}$. It can
acquire a non-vanish value, and then, gauge symmetry is
spontaneously broken (detail discussion see next subsection). In
this limit, therefore, ${\cal N}=4$ SYM reduces to a SM-like
model.
\end{enumerate}
\renewcommand{\labelenumi}{\arabic{enumi}}

\subsection*{Gauge symmetry spontaneously breaking}

In original ${\cal N}=4$ SYM, the condition of SUSY unbroken,
\begin{eqnarray}\label{3.3}
[\Phi_{AB},\Phi_{CD}]=0,
\end{eqnarray}
allows many degenerate vacuums. In particular, it allows that the
vacuum expectation value of scalar fields takes some nonzero
values. For example, we can set $<{\cal M}>=c_i T^i$ and require
at least one of real coefficients $c_i$ nonzero, where $T^i$ are
symmetric generators of $SU(N)$ group. From
decomposition~(\ref{03}) we can see that this is indeed a
supersymmetric vacuum. It is well-know as a special direction is
taken in isospin space, and gauge symmetry is spontaneously broken
partly. Consequently, part of gauge bosons acquires a mass and
others (one of parallel to this direction) still remain massless.
The former corresponds to a Higgs vacuum, and the later
corresponds to a confine vacuum. This mechanism, as showing in
many literature\cite{gauge}, reveals a method to spontaneously
break gauge symmetry but without extra Higgs bosons. In addition,
since we focus our attention on the case with $Z\to\infty$ (or
$f\to\infty$), we must expect that the expectation value $<{\cal
M}>$ (or $<\Phi_{AB}>$) is small fluctuation only and is
restricted as $Z<{\cal M}>\sim m_W$.

\subsection*{Generation and masses of fermions}

It is different from supersymmetry and conformal invariance, that
R-symmetry of ${\cal N}=4$ SYM is remained in deformation model
even when the limit $M\to\infty$ is taken. In low energy limit,
R-symmetry become generation symmetry of SM-like model, or we can
call it as general `` flavour'' symmetry. It indicates that there
should be four generation in the SM-like model. All of these
fermion are massless. However, it is because our
decomposition~(\ref{03}) is too simple. If we replace
decomposition of scalar fields $\Phi_{AB}$ by
\begin{eqnarray}\label{3.31}
(\Phi_{AB})_{ij}&=&M^{-2}e^{ia(\phi_i-\phi_j)}(\bar{\lambda}_{Aj}
\gamma_5\lambda_{Bi}-\bar{\lambda}_{Bj}\gamma_5\lambda_{Ai})
  \nonumber \\
&&-\frac{1}{2}Z_{AB}\td{U}_{ij}(\vphi)-\frac{1}{2}(1-\frac{2}{N})
 Z_{AB}\delta_{ij}, \nonumber \\
\end{eqnarray}
with $\td{U}_{ij}=e^{ia(\phi_i-\phi_j)}U_{ij}$ and real constant
$a$. Then Yukawa coupling terms of ${\cal N}=4$ SYM will yield
mass term of fermions,
\begin{eqnarray}
&&(\Phi_{AB})_{ij}[\bar{\Psi}_{Ajk}
\gamma_5\Psi_{Bki}-\bar{\Psi}_{Aki}\gamma_5\Psi_{Bjk}]\nonumber \\
&=&\left(\frac{1}{2(N-1)}+\frac{1}{N}-2\right)Z_{AB}
\bar{\lambda}_A<e^{ia(\phi_i-\phi_j)}{\cal
M}>\gamma_5\lambda_B+...,
\end{eqnarray}
when expectation value $<e^{ia(\phi_i-\phi_j)}{\cal M}>$ does not
vanish. We can see that the mass gap among different generation
is created by central charge $Z_{AB}$, and the mass gap among
different fermions in same generation is created by expectation
value $<e^{ia(\phi_i-\phi_j)}{\cal M}>$. Since $<Z{\cal M}>\sim
m_W$, the masses of heavier fermions are also order to $m_W$.

\subsection*{Toward QCD}

If $<{\cal M}>=0$, we achieve a theory with asymptotically
freedom. In other words, Lagrangian~({\ref{3.2}) exactly becomes
Lagrangian of QCD with four flavors massless fermions. Even
though gauge symmetry is spontaneously broken by a nontrivial
SUSY vacuum, the theory possesses a confine vacuum (and a Higgs
vacuum). It is difficult to separate the confine theory from
whole theory and beyond the scope of this present paper.
Alternately, there is another simple method to obtain a QCD-like
model via deformation of ${\cal N}=4$ SYM.

To replace eq.~({\ref{03}) by the following decomposition
\begin{eqnarray}\label{3.4}
\Psi_{Aij}&=&\frac{1}{\sqrt{2}}(\phi_i\lambda_{Aj}
 +\lambda_{Ai}\phi_j), \nonumber \\
\bar{\Psi}_{Aij}&=&\frac{1}{\sqrt{2}}(\phi_i\bar{\lambda}_{Aj}
+\bar{\lambda}_{Ai}\phi_j), \\
 (\Phi_{AB})_{ij}&=&M^{-2}(\bar{\lambda}_{Aj}\gamma_5\lambda_{Bi}
 -\bar{\lambda}_{Bj}\gamma_5\lambda_{Ai})
 -\frac{1}{2}Z_{AB}U_{ij}(\vphi)-\frac{1}{2}(1-\frac{2}{N})
 Z_{AB}\delta_{ij}, \nonumber
\end{eqnarray}
and impose condition $\sum_l\phi_l\lambda_{Al}=0$ as traceless
condition of spinors $\Psi_A$, we obtain a rather simple
deformation model
\begin{eqnarray}\label{3.5}
{\cal L}&=&i\sum_A\bar{\lambda}_A\;\Sla{D}\lambda_A -\frac{1}{4}
F_{\mu\nu}^aF^{a\mu\nu}+\frac{g^2\theta}{64\pi^2}
\ep^{\mu\nu\rho\sigma}F_{\mu\nu}^aF_{\rho\sigma}^a \nonumber \\
&&+\frac{1}{4} \sum_{AB}|Z_{AB}|^2Tr\left(D_\mu U^{\dag}D^{\mu}
U\right)  \nonumber \\
&&+\;\parbox{4in}{\rm four fermion terms (suppressed by $M^{-2}$
and $M^{-4}$)}
 \nonumber \\
&&+\;\parbox{4in}{\rm six fermion terms (suppressed by $M^{-6}$)}
   \nonumber \\
&&+\;\parbox{4in}{\rm eight fermion terms (suppressed by
$M^{-8}$)}.
\end{eqnarray}
It must be stressed that the condition
$\sum_l\phi_l\lambda_{Al}=0$ can be imposed only for $N>2$. The
reason is follows: In order to keep spinor $\lambda_{Al}$ as
independent dynamical degrees of freedom, not all of scalar
fields $\phi_i$ are independent, i.e., they must be solution of
equation $\sum_l\phi_l\lambda_{Al}=0$. Notice that Majorana
$\lambda_{Al}=(eQ_{Al}^*,Q_{Al})^{\rm T}$ (where $Q_{Al}$ are Weyl
spinor) have two independent components. If we want to obtain
nonzero solution of $\phi_i$ for $N=2$, equation
$\sum_l\phi_l\lambda_{Al}=0$ requires det$|Q_{A1},Q_{A2}|=0$.
Then not all of $\lambda_{Al}$ will be independent.

For $N\ge 3$, it is always possible to find some $\phi_i$ which
satisfy equation $\sum_l\phi_l\lambda_{Al}=0$ and keep all of
Majorana spinor $\lambda_{Al}$ independent. If we now take the
limit $Z\sim M\to\infty$ and $<{\cal M}>=0$, we still achieve QCD
with four massless flavors. However, we can take another special
limit: $Z=0,\;M\to\infty$. This limit also yields an theory with
asymptotically freedom and with four flavor massless fermions no
matter whether expectation value $<{\cal M}>$ vanishes or not.
This theory, in terms of Ads/CFT corresponds, can help us to
calculate low energy behaviour of QCD from QCD directly.

\section{Conclusion and Discussion}

To conclude, we propose a new algebraic deformation of ${\cal
N}=4$ SYM via decomposing fermion fields and scalar fields in
vector supermultiplets. This decomposition generates degrees of
freedom of usual quarks and leptons and a scale is introduced in
the decomposition. The deformation model is a low energy effective
(Nambu-Jona-Lasinio-like) model. In appropriate limit,
supersymmetry is broken and the deformation model reduces to a
SM-like model, or a QCD-like model. The scale, meanwhile,
corresponds to scale of supersymmetry breaking. In the SM-like
model, four generation fermions are required and the mass gaps
among these generation are created by central charge $Z_{AB}$ of
supersymmetry algebra. There are no Higgs bosons. The gauge
symmetry is spontaneously broken by nontrivial supersymmetric
vacuum. This mechanism indeed yields masses of part of gauge
bosons and mass gap among fermions of same generation. We also
show how to obtain a QCD-like model via deformation of ${\cal
N}=4$ SYM.

The essential idea of supersymmetry is symmetry between fermion
and boson. However, in traditional SUSY model we have to introduce
extra bosons or fermions as superpartner of known fermions or
bosons for achieving supersymmetry. On the contrary, in this
paper, we reveal a mechanism that SUSY can be a symmetry between
known fermions and bosons. Then it is not necessary to introduce
extra chiral supermultiplet when we construct SSM from SYM. It
also indicates that we do not need so many dynamical degrees of
freedom in any SUSY models.

There is problem on fermion mass matrix ${\cal
W}_{AB,ij}=Z_{AB}<e^{ia(\phi_i-\phi_j)}{\cal M}_{ij}>$. It is
$(4\times 4)\otimes (N\times N)$ dimension matrix and is
antisymmetric for R-symmetry indexes $A,\;B$ as well as gauge
group indexes $i,\;j$. This matrix has $2N$ real, positive and
different eigenvalues. So that every two generation fermions are
degenerated. This result is not supported by spectrum of standard
model. However, it should be pointed out that the decomposition
eqs.~(\ref{03}), (\ref{3.31}) and (\ref{3.4}) are not unique.
There are many other possibilities. For example, we can take
decomposition of Majorana spinor $\Phi_A$ as
\begin{eqnarray}\label{4.1}
\Psi_{Aij}=\frac{1}{\sqrt{2M'}}Z_{AB}\phi_i\lambda_{Bj}+....
\end{eqnarray}
This decomposition adds an extra term $\sum_C
Z_{AC}Z_{CB}\delta_{ij}/M'$ to mass matrix ${\cal W}_{AB,ij}$.
Then diagonalization on this term releases the degeneration among
those fermions and may push mass of the fourth generation fermion
to very heavy if $M'<M$. It also means that a complete study on
this type of deformation of ${\cal N}=4$ is still needed.

In this paper, our study on deformation of ${\cal N}=4$ SYM can be
easily extended to other ${\cal N}$ values. Of course, it is
unambiguous that ${\cal N}=4$ SYM is very special. It plays a
role connecting superstring theory in ten dimensions and standard
model in four dimensions. In addition, our study can be extended
to two different directions:
\begin{enumerate}
\item The idea of Ads/CFT corresponds can be used in this deformation
model. For example, an important issue is to find the
corresponding deformation of supergravity in 5-d Ads space. Then
we can calculate strong coupling limit of standard model by
Ads/CFT corresponds. In particular, it provides a feasible method
to calculate meson dynamics in low energy from QCD directly.
Another interesting issue is to study holographic renormalization
group\cite{RG} of the deformation model.
\item If we take gauge group of ${\cal N}=4$ SYM including
$SU(3)_c\times SU(2)_L\times U(1)_Y$ group, such as $SU(5)$, we
can study whether we can obtain a complete standard model via
deformation of SYM with extended supersymmetry. This deformation
may be more complicated than one suggested in this paper, but
essential idea is still that the matter fields should be
generated by decomposition of superpartner of gauge bosons.
\end{enumerate}
The studies on these aspects will be seen in future papers.

The deformation of this paper is entirely algebraic. It is not
doubted that this deformation on ${\cal N}=4$ SYM acquires
considerable success. However, we believe that this deformation
should be inspired by certain underlying dynamical mechanism.
Unfortunately, it has to still be an unsolved problem so far.

\end{document}